\documentclass[twocolumn,notitlepage,showpacs,10pt]{revtex4-2}
\usepackage{xcolor}
\usepackage[utf8]{inputenc}
\usepackage{bm}
\usepackage{graphicx}
\usepackage{tikz-cd}
\usepackage{dcolumn}
\usepackage{amsmath}
\usepackage{amssymb}
\usepackage{mathtools}
\usepackage{cancel}
\usepackage{hyperref}
\allowdisplaybreaks
\usepackage{orcidlink}
\hypersetup{citecolor=blue}
\usepackage{scalerel,stackengine}
\stackMath
\newcommand\reallywidehat[1]{%
\savestack{\tmpbox}{\stretchto{%
  \scaleto{%
    \scalerel*[\widthof{\ensuremath{#1}}]{\kern-.6pt\bigwedge\kern-.6pt}%
    {\rule[-\textheight/2]{1ex}{\textheight}}%WIDTH-LIMITED BIG WEDGE
  }{\textheight}% 
}{0.5ex}}%
\stackon[1pt]{#1}{\tmpbox}%
}
\usepackage{float}

\restylefloat{figure}

\newcommand{\dif}{\mathrm{d}}

\newcommand{\qed}{\nobreak \ifvmode \relax \else
	\ifdim\lastskip<1.5em \hskip- \lastskip
	\hskip 0.5em plus0em minus0.5em \fi \nobreak
	\vrule height0.75em width0.5em depth0.25em\fi}

\begin{document}

\title{On conformal Komar currents in LRS spacetimes}

\author{Abbas M. \surname{Sherif}\orcidlink{0000-0002-4038-6628}}
\email{abbasmsherif25@gmail.com}
\affiliation{Institute of Mathematics, Henan Academy of Sciences (HNAS), 228 Mingli Road, Zhengzhou 450046, Henan, China}

\date{\normalsize{\today}}

\begin{abstract} 
For locally rotationally symmetric (LRS) spacetimes, we construct two equivalent forms of the Komar current derived from a conformal Killing vector. One is a kinematic construction and the other is in terms of the matter quantities on the spacetime. The required conservation condition for the current is derived and discussed in various instances, and the implications of the conservation of the current, and in the case of a vanishing current, are analyzed. A relationship between the conservation criterion and the presence of trapped surfaces in the spacetime is found and discussed. We also show that for LRS II metrics with constant metric time component, the current is always conserved. In the presence of a conformal Killing horizon, properties of the current are analyzed and restrictions on, and some implications for the physical spacetime variables, in the vicinity of the horizon, are obtained. Finally, with respect to the conformal Killing horizon, the associated Noether charge is shown to be proportional to the surface gravity, establishing the thermodynamic interpretation of the Noether charge.
\end{abstract}

\maketitle

%%%%%%%%%%%%%%%%%%%%

\section{Introduction}

%%%%%%%%%%%%%%%%%%%%

The class of locally rotationally symmetric (LRS) spacetimes includes many of the well known and extensively studied exact spacetime models in general relativity \cite{el1,el2}. It  encompasses static and dynamical black hole solutions that are of both astrophysical and cosmological interest. These spacetimes admit a particular decomposition along preferred axes that allows their specification via covariantly defined variables, and the field equations to be written down as equations of first derivatives of these variables along said axes \cite{cc1,cc2}. This partial tetrad approach, in its current form, has been used, with an exceptional degree of success, in analytically understanding perturbed gravitational systems (see, for example, \cite{cc1,sem1,jp1,brad1,prat1,brad2,pluz1}). The approach has also been used to analyze properties of black hole horizons in LRS spacetimes \cite{el3,as1,as3,as4}, and more recently the evolution of these horizons in situations where the embedding spacetime is linearly perturbed \cite{as2}. The formalism has also successfully been adapted to investigating conformal symmetries in spacetimes where their existence and implications were analyzed extensively in several works \cite{gos1,gos2,as5,as6}.

The Komar current was introduced by Arthur Komar in \cite{art1,art2} and (in the particular case of relevance to this work) it is constructed from a diffeomorphism-generating vector field, the most common being those constructed from a Killing vector (KV) field. The Komar current has the property that it is divergence-free, which allows for the construction of conserved quantities like charges. Several extensions of currents and charges constructed from not necessarily exact symmetries have been developed by several authors, where their dynamics and other physical properties have been extensively analyzed. An important development in this direction includes its formulation via a covariant phase space approach by Iver and Wald \cite{wald1} (also see the work \cite{har1} by Harte which uses an affine collineation approach, as well as works of Katz {\em et al.} \cite{katz,katz1}.

Generalization of KVs goes back to seminal works of Matzner and Taubes \cite{matz1,taub} (other relatively recent works in this direction, applicable to conserved currents, include \cite{har1,huber,ruiz}). Feng proposed several constructions of conserved quantities from quite general vector fields in \cite{jfeng1}, with characterization of their properties. Some of these constructions provide a way to construct a divergence-free current from one with a non-vanishing divergence by including a gradient term, from which the charge, in the case of the Friedmann-Lemaitre-Robertson-Walker (FLRW) solution, may be assigned the notion of energy. In \cite{jfeng2}, Feng {\em et al.} introduced the Hamiltonian formulation of the so-called (1-parameter) almost-Killing equation (abbreviated AKE, whose solutions generalize Killing vectors, and includes the conformal Killing case) and studied its hyperbolicity where unboundedness of the Hamiltonian was established for all but one of the parameters. The relationship to the Komar current was considered where an exact Gauss law for a system of black holes in vacuum was exposed. Various properties and utilities of the divergence of solutions to the AKE were obtained. Chakraborty and Feng extended this analysis to the perturbative regime and established boundedness of the Hamiltonian whenever the perturbed AKE is hyperbolic \cite{jfeng3}. 

Our aim in this work is to first establish the form of a CKV-generated Komar current in a class of LRS spacetimes, and write down the condition for its conservation. These results are then tested on restrictions to certain cosmological models. We also aim to examine the criteria for a vanishing CKV-generated Komar current and comment on the restrictions these criteria impose on the region of spacetime where the CKV is defined. Finally, in some cases, CKVs generate horizons, called conformal Killing horizons (CKH), defined by the zeros of the norm of the CKV, that enclose trapped surfaces \cite{dy1,dy2}. CKV appears to play the role of the boundary that should be associated to black holes in dynamical spacetimes as with these one can define an associated Hawking temperature \cite{ted1}. We will consider constraints imposed on these horizon from the definition of the Komar current constructed from its generator. The hope is that this work may later be extended to incorporate some of the analyses carried out in the works \cite{jfeng2,jfeng3}.

This paper is structured as follows. In Section \ref{sec2}, we first introduce some relevant identities and then introduce the Komar current. We then discuss properties of the current when constructed from a conformal Killing vector. In Section \ref{sec3}, we briefly introduce LRS spacetimes. We derive the form of the Komar current constructed from a conformal Killing vector, with two alternative forms presented. One in terms of the kinematic scalars and the other in terms of the curvature variables. We then consider restrictions to the cases of certain cosmological models, and more generally, perfect fluids. The form that the conservation condition takes in certain restricted cases is also considered. Section \ref{sec4} considers the properties of conformal Killing horizons resulting from Komar current constructed from the CKV which generates the horizon, and implications for the various quantities defined on the horizon. Section \ref{sec5} concludes with a discussion of the results.

%%%%%%%%%%%%%%%%%%%%

\section{The Komar current}\label{sec2}

%%%%%%%%%%%%%%%%%%%%

We quickly review the construction of the Komar current in general relativity. We begin by writing some important relations for the purpose of this work and the benefit of the readers. We will then go on to discuss the construction and properties of the Komar current. Those already familiar with the literature may skip the first subsection.

\subsection{Useful relations}

For vectors and co-vectors the Ricci identity is given respectively by

\begin{align}
\left[\nabla_a,\nabla_b\right]Y^c={R_{bad}}^cY^d,\quad\left[\nabla_a,\nabla_b\right]Y_c={R_{abc}}^dY_d,
\end{align}
and for rank 2 tensors the Ricci identity is given by

\begin{align}
\left[\nabla_a,\nabla_b\right]Y^{cd}&=R^c_{\ fab}Y^{fd}+R^d_{\ fab}Y^{cf}.
\end{align}
The twice contracted Bianchi identity is

\begin{align}
\nabla^aR_{ab}=\frac{1}{2}\nabla_aR.
\end{align}

The commutator of the box derivative $\Box=\nabla^a\nabla_a$ and $\nabla_a$ when acting on scalars and vectors is given, respectively, by

\begin{align}
\left[\nabla_a,\Box\right]\Psi&=-R_{ab}\nabla^b\Psi,\nonumber\\
\left[\nabla_a,\Box\right]U^a&=2R_{ba}\nabla^{[a}U^{b]}+\frac{1}{2}\mathcal{L}_UR.\label{combox2}
\end{align}

In the case where the vector field $U^a$ is a CKV, i.e. $U^a$ generates the symmetry of the metric

\begin{align}
\mathcal{L}_Ug_{ab}=2\Psi g_{ab},\label{theck}
\end{align}
henceforth referred to as the {\em conformal Killing equations (CKEs)}, with $\mathcal{L}_U$ denoting the Lie derivative along $U^a$, $g_{ab}$ the spacetime metric, and $4\Psi=\nabla_aU^a$ being the conformal divergence scalar, \eqref{combox2} above is

\begin{align}
\left[\nabla_a,\Box\right]U^a=\Psi R+\frac{1}{2}\mathcal{L}_UR.
\end{align}
And a Killing vector (KV), where $\Psi=0$, would have the commutators reducing to the very simple forms

\begin{align}
\left[\nabla_a,\Box\right]\Psi=0,\quad\left[\nabla_a,\Box\right]U^a=\frac{1}{2}\mathcal{L}_UR.
\end{align}
In the case that $\Psi$ is a constant, the CKV $U^a$ is referred to as a homothetic Killing vector (HKV).

We are now in the position to introduce and discuss the current of interest.

\subsection{The current: construction and some properties}

For an arbitrary vector field $U^a$, one constructs the Komar current as

\begin{align}
J^a_K:=2\nabla_b\nabla^{[a}U^{b]},
\end{align}
with square bracket denoting anti-symmetrization on the indices, and with the obvious shift invariance for

\begin{align}
U^a\rightarrow U^a+\nabla^az,
\end{align}
for a scalar $z$.\\

Using the Ricci identity for a Rank-2 tensor, it is an easy exercise to establish that $J^a_K$ is conserved, i.e.

\begin{align}
\nabla_aJ^a_K=0.
\end{align}
For any vector $\bar{U}^a$ obeying $\nabla_a\bar{U}^a=0$, indeed the shifted current

\begin{align}
\bar{J}^a_K=J^a_K+\bar{U}^a
\end{align}
is also conserved.

For $U^a$ being a CKV, the Komar current takes the form

\begin{align}
J^a_K:=2(\nabla^a\Psi-\Box U^a),
\end{align}
with the explicit form of the divergence given by

\begin{align}
\nabla_aJ^a_K:&=2(\Box\Psi-\nabla_a\Box U^a)\nonumber\\
&=-2\left(3\Box\Psi+R\Psi+\frac{1}{2}\mathcal{L}_U R\right),
\end{align}
where we have employed the commutator of the $\Box$ derivative and the spacetime covariant derivative. The conservation of the Komar current then provides the propagation equation for the scalar curvature along $U^a$:

\begin{align}
\mathcal{L}_UR+2\Psi R+ 6\Box\Psi=0.\label{cdwe1}
\end{align}

Equation \eqref{cdwe1} is, in fact, a general property of CKVs (see, for example, \cite{jfeng1}). In other words, any $J^a_K$ constructed from a CKV in spacetime is always conserved. That is,\\

{\em For a spacetime that admits a CKV, the conformal divergence satisfies the wave equation \eqref{cdwe1}}.\\

In the case of a KV, the above is the requirement that the scalar curvature be a constant of motion along the KV. Conversely, in the case where the scalar curvature is constant along the CKV, the conservation condition is that the conformal divergence obeys the inhomogeneous wave equation.

\begin{align}
\left(\Box+\frac{1}{3}R\right)\Psi=0.
\end{align}

Now, the Komar integral can be constructed by integrating the current over some hypersurface $\Sigma$ which is equivalent to the integral of the antisymmetric part of the gradient of the generating vector field over a surface which is the boundary of $\Sigma$, by Stoke's theorem:

\begin{align}
\mathcal{Q}=\int_{\Sigma}J^a_KdS_a=\int_{\partial\Sigma}\nabla^{[a}U^{b]}\epsilon^{\partial}_{ab},\label{komint1}
\end{align}
where $dS_a$ is the volume element on the hypersurface $\Sigma$, $\partial\Sigma$ is the boundary of $\Sigma$, and $\epsilon^{\partial}_{ab}$ is the volume element on the boundary. This integral will be termed the Noether charge.

For a hypersurface that is a level surface of a function $\phi$, one may write

\begin{align}
dS_a=\sqrt{-g}\nabla_a\phi d^3y,\label{komint5}
\end{align}
where $g$ denotes the determinant of the spacetime metric and $y$ are the coordinates on hypersurface $\Sigma$, so that the Noether charge is

\begin{align}
\mathcal{Q}=\int_{\Sigma}\sqrt{-g}J^a_K\nabla_a\phi d^3y.\label{komint6}
\end{align}
This form is relevant to Section \ref{sec4} when we briefly comment on the Noether charge relative to conformal Killing horizons.

%%%%%%%%%%%%%%%%%%%%

\section{The conformal Komar current and its conservation in LRS II spacetimes}\label{sec3}

%%%%%%%%%%%%%%%%%%%%

In this section, we seek to derive the form of the Komar current in spacetimes with local rotational symmetry, constructed from a CKV. For the purpose of this work we will allow for (possibly) different causal characters of the CKV in order to facilitate a more complete consideration of the constructed Komar current.

\subsection{LRS II spacetime: a quick overview}

Locally rotationally symmetric (LRS) spacetimes admit a useful decomposition which allows for the Einstein field equations to be cast as a system of first order equations in the directional derivatives along some preferred directions. In particular, we will restrict our attention to the subclass, the LRS II class, where the preferred directions have vanishing vorticities \cite{cc1,cc2}. In local coordinates, these spacetimes are given by the metric form

\begin{align}
ds^2=-b_1dt^2+b_2dr^2+b_3d\Omega_j^2,\label{lrsmet1}
\end{align}
where $b_i=b_i(t,r)$, and $d\Omega_j^2$ is the metric on a two manifold which is flat for $j=0$, spherical for $j=1$ and a pseudo-sphere for $j=-1$. The metric \eqref{lrsmet1} admits the unit timelike and preferred spatial directions

\begin{align}
u^a=-b_1^{-1}\partial_t^a,\ n^a=b_2^{-1}\partial_r^a,
\end{align}
with $u_an^a=0$.

The interest in LRS II spacetimes is immediately obvious from the form \eqref{lrsmet1}. Many well known spacetimes, including most of the extensively studied static and dynamical spherically symmetric solutions, with or with cosmological constant, are contained in this class. To name a few we have the Schwarzschild metric, (anti)de-Sitter metric, Lemaitre-Tolman-Bondi metric, etc.

The surfaces of constant $t$ and $r$ will have induced metric

\begin{align}
N_{ab}=g_{ab}+u_au_b-n_an_b,
\end{align}
where the $g_{ab}$ is the metric on the spacetime. With this form the spacetime is said to admit a $1+1+2$ decomposition. The energy momentum tensor can now be expressed in the decomposed form

\begin{align}
T_{ab}&=\rho u_au_b+(p+\Pi)n_an_b+2Qn_{(a}n_{b)}\nonumber\\
&+\left(p-\frac{1}{2}\Pi\right)N_{ab},
\end{align}
where $\rho=T_{ab}u^au^b$ is the energy density, $3p=T_{ab}h^{ab}$ is the isotropic pressure, $Q=-T_{ab}n^au^b$ is heat flux, and $\Pi=T_{ab}n^an^b-p$ is anisotropic stress. From the Einstein equations relating $T_{ab}$ and the spacetime curvature, one can then write the Ricci tensor and scalar curvature in terms of these variables.

Derivatives along the preferred directions will be given the notations

\begin{align}
\dot{\psi}_{a\cdots b}=u^c\nabla_c\psi_{a\cdots b},\quad \psi'_{a\cdots b}=n^c\nabla_c\psi_{a\cdots b}.
\end{align}

One can then use the covariant derivatives of the preferred direction vectors to express the field equations as a system of first order partial differential equations in the dot and prime derivatives \cite{cc1,cc2}. (For LRS spacetimes the set of equations has already appeared in several works \cite{cc2,sem1,brad1,el3,gos1,as1} and we do not reproduce them here.)

Following from the decomposition, any 4-vector $\psi^a$ in a LRS spacetime may be expressed in the form

\begin{align}
\psi^a=\psi_1u^a+\psi_2n^a
\end{align}
with $\psi_1=-\psi_au^a$ and $\psi_2=\psi_an^a$ being the normal and tangential components of the vector field relative to the constant time hypersurfaces. In addition, the 2-surfaces of constant $t$ and $r$ will have the compatible covariant derivative

\begin{align}
\delta_e\psi_{a\cdots b}={N_a}^f\cdots {N_b}^g{N_e}^d\nabla_d\psi_{f\cdots g}.
\end{align}

Also, the gradient of any scalar $\psi$ on the spacetime decomposes as

\begin{align}
\nabla_a\psi=-\dot{\psi}u_a+\psi'n_a.
\end{align}

In general the dot and prime derivatives do not commute but rather obey the commutation relation, which can easily be verified for these spacetimes,

\begin{align}
(\dot{\psi})'-\left(\psi'\right)^{.}=-\mathcal{A}\dot{\psi}+\psi'u_a'n^a.\label{comrel}
\end{align}
(A caution should be exercised with the order of taking derivatives, where the derivative appearing in the parenthesis should first be taken before applying the derivative on the outside of the parenthesis.) This relation can be used to obtain constraints for exact solutions to the field equation as well as a consistency check.

With the overview out of the way, we now proceed to obtaining the Komar current from a CKV for these spacetimes.

\subsection{The current and its conservation}

The CKE \eqref{theck} may be expressed as a set of three coupled partial differential equations (PDE) for the components $\alpha$ and $\bar{\alpha}$ of the vector field \eqref{vec1} below, and a constraint equation covariantly defining the conformal divergence scalar. Existence of solutions to the CKE has been examined variously as mentioned in the introduction. When the unit directions $u^a$ and $n^a$ have non-vanishing vorticities, it was demonstrated in \cite{gos1} that the CKE always admits an HKV, which generates a horizon whenever the heat flux $Q$ attains a certain critical value. In the case that the vector field is a gradient of some differentiable function (at least $\mathcal{C}^2$), existence of solutions to the CKE was analyzed in \cite{as5} and were shown to be in bijection to the solutions of a certain wave-like PDE. (Existence in the gradient case was also covered in \cite{as6}, albeit in terms of primarily the curvature variables.) Here we derive the form of the current for the spacetimes under consideration and the conservation criterion. The succeeding subsection will give an alternative form of the current, whose properties will be used as a tool to search for CKV ``candidates'' obeying the conservation law.

According to the LRS II symmetry, we consider CKVs of the decomposed form

\begin{align}
\zeta^a=\alpha u^a + \bar{\alpha} n^a,\label{vec1}
\end{align}
where $\alpha$ and $\bar{\alpha}$ are functions of the time and radial coordinates.

We will notate $f_{ab}=\nabla_{[a}U_{b]}$, the bi-vector associated to $U^a$. Explicitly, we have \cite{as5}

\begin{align}
f_{ab}=-2\left(\dot{\bar{\alpha}}+\alpha\mathcal{A}\right)u_{[a}n_{b]}.
\end{align}

The Komar current for the associated CKV is then given by

\begin{align}
J^a_K:=-2\nabla_bf^{ab}=2\left(f u^a+\bar{f} n^a\right),
\end{align}
where

\begin{align*}
f&=-\delta_an^a\left(\dot{\bar{\alpha}}+\alpha\mathcal{A}\right)-\left(\dot{\bar{\alpha}}+\alpha\mathcal{A}\right)',\\
\bar{f}&=\delta_au^a\left(\dot{\bar{\alpha}}+\alpha\mathcal{A}\right)+\left(\dot{\bar{\alpha}}+\alpha\mathcal{A}\right)^.,
\end{align*}
with $\mathcal{A}=\dot{u}_an^a$ denoting the acceleration scalar. (The scalar $\delta_an^a$, sometimes notated $\phi$ in the literature and referred to as the sheet expansion, is the mean curvature of the constant $t$ and $r$ surfaces.) The conservation condition for the Komar current can now be expressed as

\begin{align}
\left(\dot{f}+\theta f\right)+\left(\bar{f}'+\left(\mathcal{A}+\delta_an^a\right)\bar{f}\right)=0,\label{kom21}
\end{align}
where the scalar $\theta=\mathcal{D}_au^a$ is the expansion of the timelike congruence $u^a$, with $\mathcal{D}_a$ denoting the covariant derivative along the spatial sections of the spacetime. The equation \eqref{kom21} above will then hold true for any CKV in LRS spacetime. 

Let us consider the case of an irrotational cosmological models with the metric time component $g_{tt}=-1$. Here $\mathcal{A}$ vanishes and 

\begin{align}
f=-\dot{\bar{\alpha}}\delta_an^a-\dot{\bar{\alpha}}',\ \bar{f}=\dot{\bar{\alpha}}u_a'n^a+\ddot{\bar{\alpha}},
\end{align}
so that \eqref{kom21} is now

\begin{align}
0&=\left(\ddot{\bar{\alpha}}'-\left(\dot{\bar{\alpha}}'\right)^.\right)-\dot{\bar{\alpha}}'u_a'n^a\nonumber\\
&+\left[\left(\delta_au^a\right)'-\left(\dot{\delta}_a+u_b'n^b\delta_a\right)n^a\right]\dot{\bar{\alpha}}.\label{kom22}
\end{align}

We will not provide more details here as this is quick to check, but as these are perfect fluids, the collection in the square bracket vanishes using the appropriate field equations which may be found in several works on LRS spacetimes already cited earlier. Also, using the commutation relation \eqref{comrel} on $\dot{\bar{\alpha}}$ one sees that the first two terms will cancel out. This therefore verifies the conservation criterion.

Notice that for non-accelerating metrics only the $n^a$ component of the CKV appears in the expression for the current. So, for purely temporal CKVs, the current vanishes identically ($f=\bar{f}=0$). This is the case with, for example, the FLRW with (now written in conformal coordinates)

\begin{align}
\dif s^2=a(\tau)\left(-\dif\tau^2+\frac{\dif r^2}{1-kr^2}+r^2\dif\mathcal{S}^2\right),
\end{align}
with $\dif\mathcal{S}^2$ denoting the standard metric on the unit 2-sphere, where there is the timelike CKV

\begin{align}
\zeta^a=\frac{\partial}{\partial\tau}.
\end{align}

In fact, the bivector also vanishes identically in this particular case so that the Komar integral \eqref{komint1} vanishes. (This vanishing of the Komar current and integral in the FLRW cosmological model was also argued in \cite{jfeng1}.)

In general, the vanishing of the Komar current constructed from a CKV in cosmological models with $g_{tt}=-1$ can be checked by whether or not $\bar{\alpha}$ verifies

\begin{align}
\ddot{\bar{\alpha}}+\left(\frac{1}{3}\theta+\sigma\right)\dot{\bar{\alpha}}=0,
\end{align}
obtained from the vanishing of $f$ and $\bar{f}$.

Lastly, for the sake of completeness, we will comment that the Komar integral in our case here takes the simple form

\begin{align}
-2\int_{\partial\Sigma}\left(\dot{\bar{\alpha}}+\alpha\mathcal{A}\right)u^{[a}n^{b]}\epsilon^{\partial}_{ab}.
\end{align}
If one is to consider a hypersurface $\Sigma$ such that $\partial\Sigma$ is a 2-surface with 2-metric $N_{ab}$, the integral vanishes identically. Furthermore, for a purely temporal CKV, i.e. $\bar{\alpha}=0$, were the integral to vanish over $\Sigma$, for a non-vanishing acceleration, the component $\alpha$ will change sign over $\Sigma$, i.e the hypersurface must be null somewhere.

\subsection{Alternative form of the current}

We may also write the Komar current in terms of the curvature variables. To see this, we note that the bivector $f_{ab}$ obeys

\begin{align}
\nabla_cf_{ab}=R_{abcd}\zeta^d-2\nabla_{[a}\Psi g_{b]c},
\end{align}
so that 

\begin{align}
\nabla^bf_{ba}=R_{ab}\zeta^b+3\nabla_a\Psi,\label{newfor1}
\end{align}

Now, in a LRS II spacetime the Ricci curvature decomposes as

\begin{align}
R_{ab}=a_1u_au_b+a_2h_{ab}+2Qu_{(a}n_{b)}+\Pi\left(n_an_b-\frac{}{}N_{ab}\right),
\end{align}
with the definitions

\begin{align*}
a_1=\frac{1}{2}\left(\rho+3p-2\Lambda\right),\ a_2=\frac{1}{2}\left(\rho-p+2\Lambda\right),
\end{align*}
so that the scalar curvature is $R=\rho-3p+4\Lambda$. Thus, one has the current \eqref{newfor1} as 

\begin{align}
J^a_K:=J^1_Ku^a+J^2_Kn^a,\label{cue}
\end{align}
with the definitions

\begin{align*}
J^1_K&=2\left(Q\bar{\alpha}-a_1\alpha-3\dot{\Psi}\right),\\
J^2_K&=2\left(\left(a_2+\Pi\right)\bar{\alpha}-Q\alpha+3\Psi'\right).
\end{align*}
From the equivalence of the components this imposes the constraints

\begin{align*}
2f=J^1_K,\ 2\bar{f}=J^2_K.
\end{align*}

If we denote by $k^a=u^a+n^a$ the ``outgoing'' null vector to 2-surfaces foliating the constant time hypersurfaces, we may combine the two constraints, as well as making use of the $u^au^b$ component of the conformal Killing equations given by

\begin{align}
\Psi=\dot{\alpha}+\mathcal{A}\bar{\alpha},\label{uucomp}
\end{align}
to obtain the following relation between the first derivatives of $\Psi$:

\begin{align}
\dot{\Psi}+\Psi'+\frac{1}{2}\theta_k\Psi&=-\frac{1}{2}\left(\mathcal{A}-{u_b}'n^b\right)k^a\nabla_a\alpha+\frac{1}{2}\mathcal{F}\alpha\nonumber\\
&+\frac{1}{2}\left(a_2+\Pi-Q\right)\bar{\alpha},\label{vanish0}
\end{align}
where $\theta_k=\delta_ak^a$ is the 2-divergence of $k^a$, known as the outward null expansion, and 

\begin{align*}
\mathcal{F}=\left(\theta_k+k^a\nabla_a\right)\left(\mathcal{A}-{u_b}'n^b\right)+a_1-Q.
\end{align*}
The $n^a$ component of the current is now recast as

\begin{align*}
J^2_K=3&\biggl[\left(a_2+\Pi-\frac{1}{3}Q\right)\bar{\alpha}+\left(\mathcal{F}-\frac{2}{3}Q\right)\alpha\nonumber\\
&-\theta_k\Psi-\left(\mathcal{A}-{u_b}'n^b\right)k^a\nabla_a\alpha-2\dot{\Psi}\biggr].
\end{align*}
The vanishing of the Komar current here therefore imposes

\begin{align}
0&=\theta_k\Psi+\left(\mathcal{A}-{u_b}'n^b\right)k^a\nabla_a\alpha-\left(\mathcal{F}+\frac{2}{3}(a_1-Q)\right)\alpha\nonumber\\
&-\left(a_2+\Pi-Q\right)\bar{\alpha},\label{vanish}
\end{align}
both a necessary and sufficient condition. 

As an example, consider a purely temporal CKV which aligns with $u^a$. Then, we have (where we also employ the $n^an^b$ component of the conformal Killing equations given by $\Psi=\alpha{u_b}'n^b$)

\begin{align}
0=\left(\theta_k+\mathcal{A}-{u_b}'n^b-\left(\mathcal{F}+\frac{2}{3}(a_1-Q)\right)({u_b}'n^b)^{-1}\right)\Psi.\label{vanish1}
\end{align}
That is, either the CKV is a KV or

\begin{align}
-\theta_k=\mathcal{A}-{u_b}'n^b-\left(\mathcal{F}+\frac{2}{3}(a_1-Q)\right)({u_b}'n^b)^{-1}.\label{vanish2}
\end{align}

Now, the quantity $\theta_k$ plays a very fundamental role in the study of the evolution of black holes (see the references \cite{ash1,booth4}, and the for the literature relevant to LRS spacetimes and the formalism employed here, see \cite{el3,as1,as3,as4}). If at all points of a surface $\mathcal{S}$, $\theta_k=0$, $\mathcal{S}$ is called a marginally outer trapped surface (MOTS). These surfaces foliate horizons of black holes. For example, the Schwarzschild event horizon as well as dynamical horizons in the Lemaitre-Tolman-Bondi and the Vaidya (both charged and neutral cases) spacetimes are MOTS. (We emphasize here that MOTS need not foliate horizons enclosing black holes, for example those in the (anti)de-Sitter and Robertson-Walker spacetimes.) We note that the equation \eqref{vanish} allows for an interesting formulation of the MOTS condition:\\

{\em Let $M$ be a LRS II spacetime with vanishing Komar current constructed from a (proper) CKV \eqref{vec1}. Then, any surface $\mathcal{S}\subset M$ on which}

\begin{align}
0&=\left(\mathcal{A}-{u_b}'n^b\right)k^a\nabla_a\alpha-\left(\mathcal{F}_{\theta_k=0}+\frac{2}{3}(a_1-Q)\right)\alpha\nonumber\\
&-\frac{8}{3}\left(a_2+\Pi-Q\right)\bar{\alpha}\label{vanish3}
\end{align}
{\em is a MOTS.}\\

The ``proper'' here indicates that $\Psi$ is not a constant. In the case of a pure temporal vector field this MOTS condition is simply

\begin{align}
0=\mathcal{A}-{u_b}'n^b-\left(\mathcal{F}_{\theta_k=0}+\frac{2}{3}(a_1-Q)\right)({u_b}'n^b)^{-1}.\label{vanish4}
\end{align}
So, in the case considered earlier with a perfect LRS II fluid with vanishing acceleration, \eqref{vanish4} reduces to

\begin{align}
({u_b}'n^b)^2+\mathcal{F}_{\theta_k=0}+\frac{2}{3}a_1=0,\label{vanish5}
\end{align}
thereby imposing

\begin{align}
\mathcal{F}_{\theta_k=0}+\frac{2}{3}a_1=\frac{5}{3}a_1-k^a\nabla_a({u_b}'n^b)<0.\label{vanish6}
\end{align}
This is an upper bound on the local energy density. Thus, in these spacetimes, these MOTS will exist only in regions obeying the inequality \eqref{vanish6}.

We may also compare the vanishing current condition \eqref{vanish} with \eqref{vanish0} to obtain the first order PDE

\begin{align}
\dot{\Psi}+\Psi'=\frac{1}{3}(a_1-Q)\alpha.\label{alp}
\end{align}
This equation provides certain restrictions on HKVs with vanishing current: for any perfect LRS II fluid, an HKV generating a vanishing Komar current either points along the radial direction, or $a_1$ vanishes. In the case of vacuum this is the condition of a vanishing cosmological constant. Let us note now that for \eqref{alp} no matter type has been imposed and hence it is general for the LRS II class of spacetimes.

For the conservation condition in the $J_K^i$ components, we comment only on a restricted case of certain LRS II solutions with perfect fluid-type matter. For a perfect fluid-type matter, $Q=\Pi=0$. Here, consider vacuum solutions. Then, we have

\begin{align*}
J^1_K=2\left(\Lambda\alpha-3\dot{\Psi}\right),\quad J^2_K=2\left(\Lambda\bar{\alpha}+3\Psi'\right).
\end{align*}
Clearly, even for an HKV (including a KV), in this case, the current will vanish if and only if we have a vanishing $\Lambda$. That is, $\Lambda$-vacuum LRS II solutions with conformal symmetry will have a non-trivial Komar current. By using \eqref{alp} it follows that that the vanishing of the current imposes

\begin{align}
(\Lambda-a_1)\alpha+\Lambda\bar{\alpha}=0.
\end{align}
If $\Lambda$ vanishes, then either $\zeta^a$ points along the radial direction or the strong energy condition must marginally hold: $\rho+3p=0$.

The conservation condition now reads

\begin{align}
0&=\dot{J}_K^1+{J_K^2}'+\theta J_K^1+\left(\mathcal{A}+\delta_an^a\right)J_K^2\nonumber\\
&=-\ddot{\Psi}+\Psi''-\theta\dot{\Psi}+\left(\mathcal{A}+\delta_an^a\right)\Psi'+\Lambda\Psi.\label{conseh1}
\end{align}
In compact form, this PDE is the operator equation

\begin{align}
\left(\Box-\Lambda\right)\Psi=0.\label{conseh2}
\end{align}
That is, the Komar current constructed for a CKV \eqref{vec1}, under said restrictions, is conserved if an only if the above homogeneous PDE is satisfied. This condition is generic for CKV, as previously noted. Indeed, it follows that an HKV is not possible for these spacetimes with non-vanishing $\Lambda$. Furthermore, equation \eqref{conseh2} implies that the Komar current constructed from any KV is conserved. 

And since for vacuum the conformal divergence $\Psi$ obeys the homogeneous wave equation, it follows that the cosmological constant will not vanish only if $\zeta^a$ is a KV. 

For the case of an HKV (and not a KV, in which case we must have a vanishing $\Lambda$), if $\alpha\neq0$, we must have $Q=0$ and $a_1=0$ (in this case the strong energy condition is marginally satisfied). Impose that the current must vanish. Then, either $\bar{\alpha}=0$ in which case the vector field points along $u^a$, or the following particularly restrictive condition holds:

\begin{align}
\frac{2}{3}\rho+\Pi=0.
\end{align}
It is clear from the above equation that any perfect LRS II fluid in this case will have to be a vacuum solution. The KV case is similarly analyzed, except that in the case of the former the spacetime has to be time symmetric, i.e. one must have vanishing shear and expansion.

On the other hand, if one imposes that the strong energy condition strictly holds, then $\alpha=0$. Specialize to a perfect fluid, then, the vanishing condition gives

\begin{align}
\theta_k\bar{\psi}-2a_2b=0,
\end{align}
which upon using the $u^au^b$ component of the conformal Killing equations we have

\begin{align}
\theta_k\mathcal{A}=a_2,
\end{align}
a constraint providing yet another definition of a MOTS: let $\zeta^a$ be an HKV in a perfect LRS II fluid with vanishing cosmological constant, and let $\mathcal{U}$ be a region of the spacetime on which the associated Komar current vanishes. Then, any surface $\mathcal{S}\subset\mathcal{U}$ on which the local energy density and pressure coincide is a MOTS.

Finally, we remark that there are several ways that one may use the derived equations to search for a conformal Killing vector and hence define its associated Komar current. Two approaches are described below which look for CKV candidates. 

\begin{enumerate}

\item {\em A mode of construction 1.:} One way to construct a Komar current generated by a CKV-like vector field as have been analyzed is as follows. Begin by finding a solution $\bar{\psi}$ to \eqref{conseh2}. Then, if the solution $\bar{\psi}$ factors as

\begin{align*}
\bar{\psi}= a\delta_au^a+b\delta_an^a,
\end{align*}
for some functions $a(t,r)$ and $b(t,r)$, the Komar current \eqref{cue}, now with

\begin{align}
J_K^1(\bar{\psi})&=4\left(Qb-a_1a-\frac{3}{2}\dot{\bar{\psi}}\right),\nonumber\\
J_K^2(\bar{\psi})&=4\left((a_2+\Pi)b-Qa+\frac{3}{2}\bar{\psi}'\right),
\end{align}
is constructed from the vector field 

\begin{align}
\frac{1}{2}\zeta^a=au^a+bn^a,\label{impo1}
\end{align}
which is a CKV candidate.

In order to search for functions of the form \eqref{impo1}, one may look at a combination of the field equations to obtain functions of this form. After substituting the corresponding combination into \eqref{conseh2} and expanding out, impose the necessary conditions for the resulting equation to hold.

Perhaps the approach outlined above is not necessarily efficient. However, it does avoid directly solving the conformal Killing equations, which helps to reduce the computational expense of the procedure.

\item {\em A mode of construction 2.:} Let $q$ be some function on the spacetime with a non-vanishing acceleration and suppose a function $\bar{\psi}$ solves

\begin{align}
\dot{\bar{\psi}}+\bar{\psi}'=(a_1-Q)q.
\end{align}
Then, the vector field

\begin{align}
\frac{1}{3}\zeta^a=qu^a+\left(\frac{\bar{\psi}-\dot{q}}{\mathcal{A}}\right)n^a,
\end{align}
is a CKV candidate which generates the current \eqref{cue}, but now with

\begin{align}
J_K^1(\bar{\psi})&=6\left(Q\left(\frac{\bar{\psi}-\dot{q}}{\mathcal{A}}\right)-a_1q-\dot{\bar{\psi}}\right),\nonumber\\
J_K^2(\bar{\psi})&=6\left((a_2+\Pi)\left(\frac{\bar{\psi}-\dot{q}}{\mathcal{A}}\right)-Qq+\bar{\psi}'\right).
\end{align}

\end{enumerate}

It is worth commenting that in certain cases where the spacetime has some additional symmetry or restricted kinematics, additional conditions to ensure that the above vector field is indeed a CKV may be specified or verified. A simple example is the case of a purely temporal CKV for a shear-free solution. Consider the first construction. Then, for the CKV candidate \eqref{impo1} to be a CKV, it is sufficient that $a$ verifies

\begin{align}
a'=\mathcal{A}a.
\end{align}
This is easily seen by noting that

\begin{align}
\nabla_a\zeta_b=\bar{\psi}g_{ab}+f_{ba}\iff \eqref{theck}.\label{ckenew}
\end{align}
Expanding the gradient of \eqref{impo1} it can be checked that the left hand side of the above is fulfilled if and only if  the following hold:

\begin{align}
\sigma a=0;\quad a'=\mathcal{A}a,
\end{align}
with the former obviously being satisfied in the shear-free case. Here $\sigma$ may be obtained by projecting the shear tensor $\sigma_{ab}=\nabla_{(a}u_{b)}$ to the 2-surfaces foliating spacelike hypersurfaces. Similar simplifications are possible in other cases.

%%%%%%%%%%%%%%%%%%%%

\section{On conformal Killing horizons and conserved currents from their generators in LRS II spacetimes}\label{sec4}

%%%%%%%%%%%%%%%%%%%%

We now consider the Komar current in the presence of a conformal Killing horizon (CKH) \cite{dy1,dy2} (the rotating counterpart to these horizons was introduced in \cite{gh1}). We take this hypersurface to be the one over which the Komar integral is evaluated. We discuss those situations where the current vanishes on the CKH and what the conservation equation looks like. Before moving on to these considerations, let us first derive some additional properties of the CKH.

A CKH is a hypersurface on which the norm of a timelike conformal Killing vector field vanishes \cite{dy1}. Specifically, we will be interested in those conformal Killing vector fields for which the gradient of its divergence vanishes nowhere on the CKH (shortly we will emphasize why this consideration is imposed). 

In \cite{as4}, the necessary and sufficient condition for the existence of a conformal Killing horizon (CKH) in LRS spacetimes was derived. In particular, one requires that the components of the generator of the horizon be constant along the generator itself:

\begin{align}
0=\mathcal{L}_\zeta\alpha=\mathcal{L}_\zeta\bar{\alpha},
\end{align}
with neither the components $\alpha$ nor $\bar{\alpha}$ being constant. Furthermore, it is imposed that there are no homothetic points on the CKH, i.e. no $p$ on the CKH for which $D_a\Psi\overset{p}{=}0$, where $D_a$ denotes the compatible covariant derivative on the CKH.

By projecting \eqref{theck} with $N^{ab}$ the form of the conformal divergence can be expressed as

\begin{align}
\Psi=\frac{1}{2}\alpha\delta_au^a+\frac{1}{2}\bar{\alpha}\delta_an^a.\label{conevo1}
\end{align}
Taking the Lie derivative of \eqref{conevo1} along the CKV, upon evaluation at the CKH, gives 

\begin{align}
\mathcal{L}_{\zeta}\Psi=\frac{1}{2}\alpha\mathcal{L}_{\zeta}\delta_au^a+\frac{1}{2}\bar{\alpha}\mathcal{L}_{\zeta}\delta_an^a.\label{conevo2}
\end{align}
And since we have that 

\begin{align}
\mathcal{L}_{\zeta}\Psi=\alpha\dot{\Psi}+\bar{\alpha}\Psi',\label{conevo3}
\end{align}
the value of the respective derivatives at the CKH can be evaluated as

\begin{align}
\dot{\Psi}&=\frac{1}{2}\mathcal{L}_{\zeta}\delta_au^a,\label{conevo4}\\
\Psi'&=\frac{1}{2}\mathcal{L}_{\zeta}\delta_an^a.\label{conevo5}
\end{align}

Associated with the horizon, one may define an invariant surface gravity \cite{ted1} given by

\begin{align}
-2\kappa\zeta_a=\nabla_a(\zeta_b\zeta^b),
\end{align}
which is constant on the horizon. It can indeed be shown, using properties of $\zeta^a$ as a CKV, that \cite{as4}

\begin{align}
-\zeta_a\mathcal{L}_{\zeta}\kappa=\zeta_b\zeta^b\nabla_a\Psi.
\end{align}
If $\nabla_a\Psi$ vanishes nowhere that $\zeta^a$ is defined, $\kappa$ is constant along the CKH. The consideration in \cite{as4} was providing a necessary and sufficient condition for the existence of a conformal Killing horizon which requires the derivative $\mathcal{L}_{\zeta}\kappa$ vanishing on it. Where $\nabla_a\Psi$ vanishes, $\kappa$ is constant there and does not necessarily characterize the vanishing of the norm of $\zeta^a$ at that point. For this reason, it was imposed in \cite{as4} (and therefore the condition is imposed here for the purpose of this work) that the gradient of $\Psi$ is non-vanishing.

Now, the constancy of the surface gravity may be expressed as \cite{as4}

\begin{align}
\mathcal{L}_{\zeta}\Psi=\mathcal{L}_{\zeta}\left(\dot{\bar{\alpha}}+\mathcal{A}\alpha\right).\label{conevo6}
\end{align}
The Lie derivative of \eqref{uucomp} along $\zeta^a$, equated to \eqref{conevo6}, implies that on the CKH we must have

\begin{align}
\mathcal{L}_{\zeta}\left(\dot{\alpha}+\dot{\bar{\alpha}}\right)=0.\label{conevo7}
\end{align}
We note this also implies that on the horizon  one has that

\begin{align}
\mathcal{L}_{\zeta}\left(\alpha'+\bar{\alpha}'\right)=0.\label{conevo7}
\end{align}

Now, another result of interest was obtained for LRS spacetimes (here we restrict to the LRS II class) with non-vanishing $\theta$. It was demonstrated in \cite{as4} that for any timelike vector field of the form \eqref{vec1}, the condition

\begin{align}
\dot{\bar{\alpha}}+\frac{1}{2}\alpha\delta_an^a=0,\label{gua1}
\end{align}
guarantees that the vector field is a CKV and necessarily generates a CKH. The Lie derivative of \eqref{gua1} along $\zeta^a$, using \eqref{conevo5} and \eqref{conevo7}, gives

\begin{align}
\mathcal{L}_{\zeta}\dot{\alpha}-\alpha\Psi'=0.\label{conevo8}
\end{align}

Also, by the constancy of $\kappa$ it is also found that on the CKH,  one has $\mathcal{L}_{\zeta}\left(\dot{\alpha}-\dot{\bar{\alpha}}\right)=0$ \cite{as4}, so that $\mathcal{L}_{\zeta}\dot{\alpha}=\mathcal{L}_{\zeta}\dot{\bar{\alpha}}=0$. Thus, by \eqref{conevo8}, along the CKH, $\Psi'=0$, i.e. the conformal divergence is constant along integral curves of $n^a$, along the CKH. (Note that the gradient $\nabla_a\Psi$ is then always timelike.) This then says that the sheet expansion $\delta_an^a$ is necessarily constant on a CKH in an expanding LRS II spacetime, by \eqref{conevo5}. 

Another consequence of the vanishing of $\Psi'$ is then the following result for LRS II solutions with perfect fluid type matter:\\

{\em Consider a CKH in a perfect LRS II fluid with non-vanishing expansion and a vanishing cosmological constant, generated by a CKV $\zeta^a$ with non-constant components. Then, the Komar current associated to $\zeta^a$, along the CKH generated by $\zeta^a$, is future-directed.}\\

What the above says is that the usual timelike observers in these spacetimes will experience the current flowing through a CKH. It is also indeed clear that the Komar current is non-vanishing on the CKH as otherwise this would lead to homothetic points on the horizon. The vanishing of the Komar current in the case of the Robertson-Walker metric therefore verifies the non-existence of such horizons.

Notice from the $J_K^2$ component of the current that a necessary condition that the current vanishes on the CKH is that 

\begin{align}
\left(a_2+\Pi\right)\bar{\alpha}-Q\alpha=0.\label{conevo10}
\end{align}
And since on the CKH the components of the CKV coincide, one has the above reducing to the following condition on the CKH $a_2+\Pi-Q=0$. So, we see that for perfect LRS II fluids with vanishing cosmological constant, the spacetime Ricci tensor restricted to the CKH is

\begin{align}
R_{ab}\overset{CKH}{=}a_1u_au_b,
\end{align}
i.e. $u^a$ is a Ricci eigenvector. That is, there is an additional restriction on the spacetime Ricci tensor evaluated to the horizon. The component $a_1$ of the Ricci tensor specifically obeys the constraint

\begin{align}
-\frac{2}{3}a_1=k^a\nabla_a\delta_bu^b.
\end{align}
That is, the restriction of the Ricci tensor to the CKH is specified by the ``evolution'' of the 2-divergence of the timelike congruences along outgoing null geodesics. 

We now quickly comment on a property of a CKH that is independent of the definition of the current. As can be seen, on the CKH with $\alpha=\bar{\alpha}$, \eqref{conevo2} can now be written as 

\begin{align}
\mathcal{L}_{\zeta}\Psi=\frac{1}{2}\alpha\mathcal{L}_{\zeta}\theta_k,\label{mots1}
\end{align}
But as $\Psi'$ vanishes on the CKH, we have that $2\dot{\Psi}=\mathcal{L}_{\zeta}\theta_k$. So, the expansion $\theta_k$ cannot have a vanishing gradient at any point along the conformal orbits, i.e. $\mathcal{L}_{\zeta}\theta_k\neq0$, since this would otherwise imply $\dot{\Psi}=0$ there, i.e. $\Psi$ has a homothetic point which was ruled out for our current consideration. On the other hand, if any of the surfaces lying in the CKH is a MOTS of constant $t$ and $r$, it follows that $\mathcal{L}_{\zeta}\theta_k=0$, and we draw the same conclusion as before. Thus, for our particular consideration in this work, the CKH cannot be foliated by MOTS. That is, 

\begin{align}
\zeta_a\zeta^a\not\propto\theta_k.
\end{align}

Finally, we comment on the integral \eqref{komint6}. Taken over the CKH, the integral \eqref{komint6} is found to take the form (we recall that the surface gravity $\kappa$ is constant along the CKH)

\begin{align}
\mathcal{Q}=-2\kappa\int_{CKH}\left(\alpha J^1_K+\bar{\alpha}J^2_K\right)\sqrt{-g}d^3y,
\end{align}
which is explicitly given by

\begin{align}
\mathcal{Q}=-4\kappa\int_{CKH}\left(Q\bar{\alpha}^2-a_1\alpha^2-3\alpha\dot{\Psi}\right)\sqrt{-g}d^3y,
\end{align}
which upon using \eqref{alp} becomes

\begin{align}
\mathcal{Q}=-4\kappa\int_{CKH}\left(Q\bar{\alpha}^2-(2a_1-Q)\alpha^2\right)\sqrt{-g}d^3y.
\end{align}

The thermodynamic interpretation of $\mathcal{Q}$ is now therefore explicit, due to the presence of the surface gravity $\kappa$. And as it is seen, if there is no flux of matter, i.e. $Q=0$, the charge is computed as in the case of a purely temporal generator, with

\begin{align}
\mathcal{Q}=8\kappa\int_{CKH}a_1\alpha^2\sqrt{-g}d^3y.
\end{align}
We know that $\alpha$ can be zero nowhere along the CKH as at such a point the conformal divergence would be constant along $\zeta^a$. Therefore, if the strong energy condition strictly holds across the CKH, the CKH will have a non-vanishing Noether charge. In fact, by the restriction that the conformal divergence is nowhere constant, it is also non-constant along null geodesics. Thus, \eqref{alp} ensures that the strong energy condition necessarily strictly holds if we simply require that the strong energy condition holds.

%%%%%%%%%%%%%%%%%%%%

\section{Concluding remarks}\label{sec5}

%%%%%%%%%%%%%%%%%%%%

The role of symmetries in theoretical physics is extensive, where their imprints can be found in various sectors of general relativity and quantum mechanics. In general relativity, a prevalent usage of symmetries is the generation of new exact solutions from existing ones that admit conformal symmetries. Yet, another well known application of symmetries is the construction of conserved quantities in spacetimes. One such quantity is the Komar current (sometimes referred to as the Noether current), from which one may construct a charge that may be associated to mass and other physical observables, by integrating the current over some hypersurface.

This work has examined the construction of the Komar current and the Noether charge in the class of locally rotationally symmetric spacetimes. Owing to the symmetry of these spacetimes, we have employed a well-adapted covariant formulation and constructed two equivalent forms of the Komar current: a kinematic form and a geometric form. This equivalence is exploited to obtain an inhomogeneous first order partial differential equation for the divergence of the CKV, which we called the conformal divergence. This equation may be used to generate a CKV and consequently define a Komar current. The condition for conservation of the current, as well as its vanishing, is discussed, from which several results related to restrictions on some of the spacetime variables are obtained. The relationship of properties of the current to the trapping of surfaces of constant time and radius are also obtained and discussed. That is, through the definition of current, we are able to possibly detect the presence of black holes from their horizon cross sections, or marginally outer trapped surfaces (MOTS).

Finally, we have considered conformal Killing horizons (CKH), null hypersurfaces on which the norm of a CKV vanishes, which admit no homothetic points (those points on the CKH where the gradient of the conformal divergence vanishes). Several additional properties of the CKH are derived in the case that the constant time hypersurfaces have non-vanishing expansion. It is shown that on a CKH in LRS II spacetimes, the conformal divergence will be constant in the radial direction, which consequently requires that the expansion of the constant time and radius surfaces, called the sheet expansion, is also constant along the CKH. These properties rule out that the CKH can be foliated by MOTS. The form of the Noether charge along the CKH is then computed and is found proportional to the invariant surface gravity associated to the CKH. For perfect fluid solutions, it is shown the the strong energy condition being valid along the CKH ensures that the CKH has a non-vanishing Noether charge.

%%%%%%%%%%%%%%%%%%%%

\section*{Acknowledgement}

%%%%%%%%%%%%%%%%%%%%

It is acknowledged that an anonymous referee's helpful comments have added a degree of clarity to some of the claims in the manuscript. The author acknowledges that this research is supported by the Institute of Mathematics, funded through the High-level Talent Research Start-up Project Funding of the Henan Academy of Sciences (Project No.: 251819085).

\end{document}